\documentclass[a4paper,11pt]{article}
\usepackage{pos}

\begin{document}

\title{A brief overview of the baryon cycle in galaxies}

\author*[a]{S. Veronese}

\affiliation[a]{Max Planck Institute for Radio Astronomy, Auf dem Hügel 69, 53121 Bonn, Germany}

\emailAdd{veronese.astro@gmail.com}

\abstract{Galaxy formation and evolution is governed by the continuous exchange of gas between the cosmic web, the circumgalactic medium (CGM), and the interstellar medium (ISM) of galaxies. This brief review summarises our current understanding of this baryon cycle, from the accretion of gas onto galaxies to its conversion into stars, and the feedback processes that regulate the growth of galaxies. Several open questions will be highlighted, that will encourage progresses in cosmological simulations, multiwavelength observations, and theory to resolve.}

\FullConference{Mondello Workshop 2026 Frontier Research in Astrophysics – V (FRAPWS2026)\\
15-20 June 2026\\
Palermo, Italy\\}

\maketitle

\section{Introduction}\label{sec:intro}
In the $\Lambda$ Cold Dark Matter ($\Lambda$CDM) paradigm (see reviews by \citep{perivola22,primack24}), the gravitational collapse of baryons within the initial distribution of dark matter haloes led to the formation of a complex structure of voids, filaments, and nodes known as the cosmic web, ubiquitous in the observed Universe (see, e.g., \citep{springel06,zavala19} for a brief overview, \citep{tempel14,jasche15,darvish17} for empirical evidence, and \citep{springel05b,dalal08,cautun14} for theoretical predictions). Galaxies are assembled primarily within filaments, whereas voids are the most underdense regions of the Universe, and nodes host galaxy clusters. Further growth of galaxies is regulated by the interplay between gas accretion from the cosmic web to the circumgalactic medium (CGM) and down to the interstellar medium (ISM) of galaxies (reviewed by \citep{sancisi08,putman12,tumlinson17,peroux20,faucher23}) and feedback from stellar activity \citep{zhang18,tacconi20,thompson24} and active galactic nuclei (AGN; \citep{fabian12,heckman14,harrison24}).
\\\indent A comprehensive theoretical and observational framework for galaxy formation and evolution therefore requires connecting physical phenomena spanning approximately ten orders of magnitude in spatial scale and several orders of magnitude in time-scale. Galaxy assembly within the cosmic web occurs on Gpc-to-Mpc scales over Gyr time-scales, and gas accretion proceeds on Mpc-to-kpc scales, again over Gyr time-scales. Once inside the galaxy, gas cools and settles dynamically on kpc-to-pc scales over Myr time-scales, before star formation and AGN activity take place on pc-to-km scales over time-scales as short as a Myr.
\\\indent The multi-scale nature of galaxy formation and evolution is not the only challenge. The baryon cycle \citep{peroux20}, which describes how gas accretes onto galaxies, cools to form stars, and is partially returned to the CGM through stellar- and AGN-driven outflows, depends on galaxy mass \citep{scholte24}, galactic environment \citep{boselli06}, and cosmic time \citep{forster20,walter20}. This renders galaxy formation and evolution not only an inherently multi-scale problem, but a complex dynamical puzzle that requires input from cosmological simulations, multiwavelength observations, and analytic theory in equal measure.
\\\indent This brief review covers several aspects of the baryon cycle. Section\,\ref{sec:accretion} discusses gas accretion from the CGM to the ISM; the process of star formation is presented in Sect.\,\ref{sec:sf}, and the impact of stellar feedback on the baryon cycle is described in Sect.\,\ref{sec:stellarfeedback}; the feeding of supermassive black holes (SMBHs) at the centres of galaxies and the contribution of AGN outflows to the baryon cycle are discussed in Sects\,\ref{sec:agnfeeding} and \ref{sec:agnfeedback}, respectively. Concluding remarks, including a brief summary of the open questions, are given in Sect.\,\ref{sec:conc}.

\begin{figure}[!t]
    \centering
    \includegraphics[width=\linewidth]{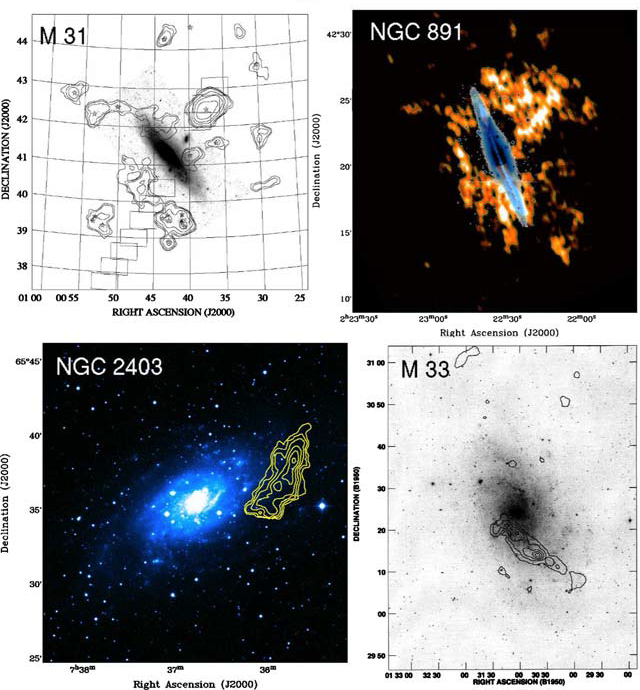}
    \caption{Examples of H\,{\sc i} complexes thought to trace accretion from the CGM and/or the cosmic web. The top two panels show gas clouds around M31 and NGC 891, whereas the bottom panels present the filamentary structures in NGC 2403 and M33. Image credit: \citep{sancisi08}.}
    \label{fig:accretion}
\end{figure}

\section{Gas Accretion}\label{sec:accretion}
The accretion of pristine gas from the cosmic web and processed gas from the CGM is invoked to explain the observed cosmic star formation rate (SFR) density \citep{madau14,walter20}. Indeed, the cosmic SFR density rises steeply to $z\sim2$ and declines thereafter, while the gas depletion time inferred from CO observations is comparatively short ($\sim$Gyr) and remains roughly constant with redshift \citep{saintonge13,saintonge22}. Without continued replenishment of gas, galaxies should have exhausted their gas reservoirs and shut off star formation many Gyr ago. However, the precise mechanisms by which this gas is delivered to galaxies remain an open question from both observational and theoretical perspectives.
\\\indent From a theoretical standpoint, gas accretion is expected to depend on halo mass. For dark matter halo masses below $10^{8{\rm-9}}$ M$_\odot$, the density of the inflowing gas is lower than the self-shielding density, so the gas is heated by photoionisation from the cosmic ionising background and accretes inefficiently \citep{noh14}. In more massive haloes, the gas is shock-heated to temperatures above $10^5$ K. When the cooling time is shorter than the free-fall time, accretion proceeds mainly in a cold mode, in which streams and clouds of cool ($T<10^5$ K) gas flow directly onto galaxies along cosmic filaments \citep{keres05,keres09,dekel09,dekel09b,voort11,danovich15,nelson16}. In contrast, when the cooling time exceeds the free-fall time, gas accretion occurs in a hot mode, whereby the gas forms a stable hot CGM rather than immediately fuelling star formation. Over time, this hot CGM can still accrete onto the galaxy through cooling flows \citep{thompson16,schneider18,schneider20,fielding22,hafen22}, thermal instabilities \citep{mccourt12,sharma12}, and star-formation-induced galactic fountains \citep{fraternali13,fraternali17}.
\\\indent While direct detection of hot accretion flows is difficult owing to the low column density and high ionisation energy of the gas, cold gas inflows from the CGM have been observed at $z>2$ \citep{heintz25,bevacqua26}, suggesting that cold-mode accretion dominated in the early Universe, as predicted by simulations \citep{correa18,waterval25}. At $z\sim0$, instead, the situation is less clear. Mergers and tidally stripped gas account for less than $\sim20\%$ of the gas accreted onto star-forming galaxies \citep{diteodoro14}, with the remaining $\gtrsim80\%$ presumed to originate from the CGM. But observations of cold gas around nearby galaxies, probed via neutral atomic hydrogen (H\,{\sc i}) emission, have revealed kinematically anomalous gas \citep{fraternali02,braun04,pisano14,oosterloo07,wolfe16,xu22,liu23,das24}, whose origin remains unclear (see Fig.\,\ref{fig:accretion}), even with the most recent deep H\,{\sc i} surveys \citep{mhongoose2,veronese25,yang25,wang26}. Furthermore, the estimated accretion rate at $z\sim0$ does not match the measured SFR in galaxies \citep{kamphuis22} and a detailed comparison between the H\,{\sc i} around simulated and observed galaxies at $z\sim0$ revealed a remarkable tension not only in the gas distribution but also in the gas dynamics \citep{marasco25}.

\begin{figure}[!t]
    \centering
    \includegraphics[width=\linewidth]{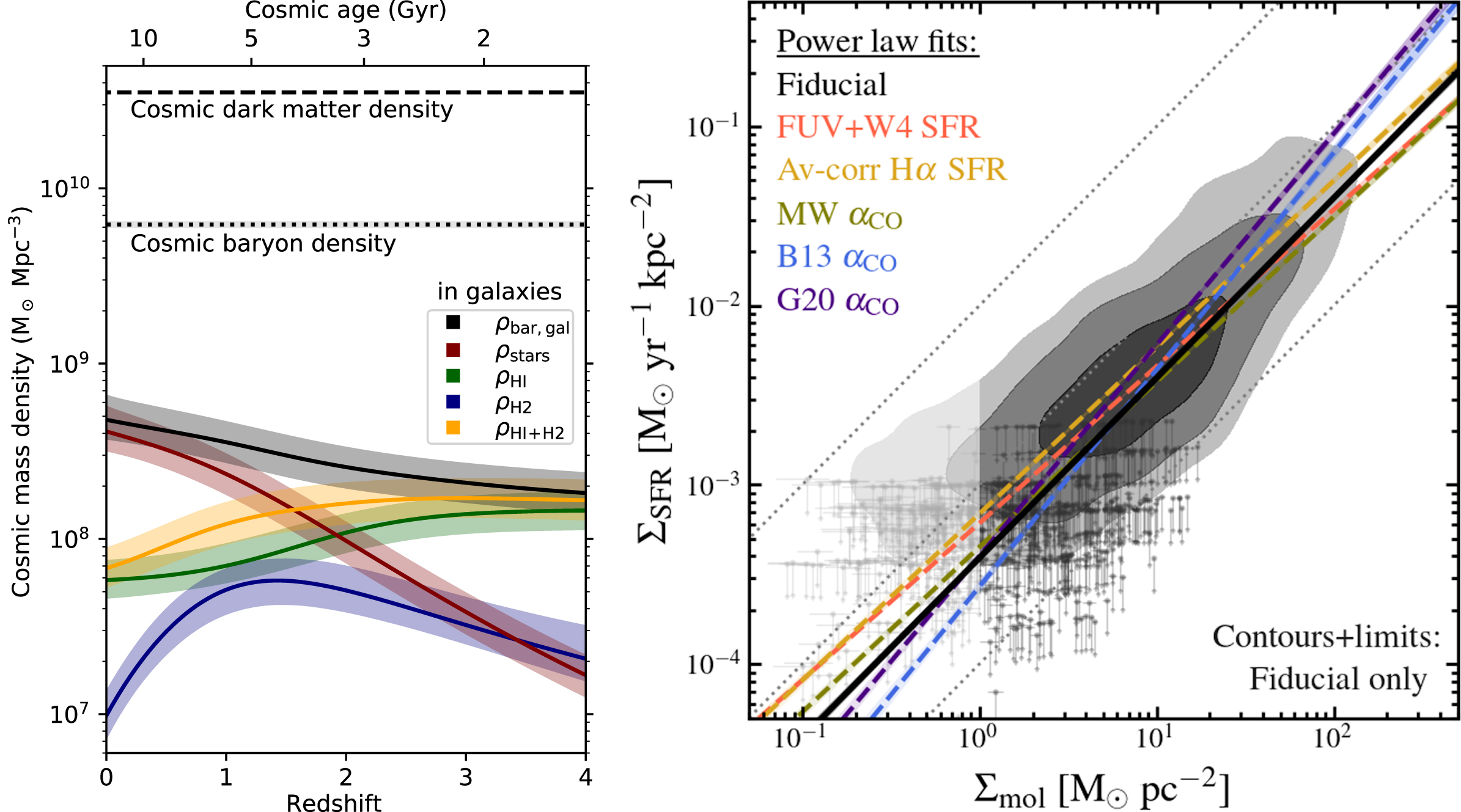}
    \caption{The left panel shows the redshift evolution of different baryonic components (baryons associated with galaxies in black, stars in maroon, H\,{\sc i} in green, H$_2$ in blue, and gas in orange) as presented by \citep{walter20}. For each component the solid line is the best fit and the shaded region the $1\sigma$ uncertainty. The right panel shows the Kennicutt--Schmidt relation from \citep{sun23}. The density contours at the 40\%, 80\%, and 95\% levels denote the spatially resolved regions of the galaxy sample for which both the SFR and molecular gas content were measured. The downward arrows correspond to $3\sigma$ upper limits. The solid black line shows the fiducial best-fit power-law model, while the other coloured lines refer to best fits obtained using alternative SFR calibrations or molecular gas conversion factors. The thin dotted lines correspond to linear relations assuming constant molecular gas depletion times of 0.1, 1, and 10 Gyr (top to bottom).}
    \label{fig:sf}
\end{figure}

\section{Star formation}\label{sec:sf}
The accreting gas from the cosmic web and the CGM settle in a rotating disc and form the ISM. The ISM comprises distinct gas phases: hot ($T\sim10^4\,$K) ionised gas \citep{haffner09,chevance20,belfiore22}, warm ($T\sim10^2$--$10^4\,$K) neutral gas \citep{wolfire95,wolfire03,mcclure23,pingel24,blok26}, cool ($T\sim10^2\,$K) neutral gas \citep{field69,kanekar11}, and cold ($T<10^2\,$K) molecular gas \citep{young91,saintonge17,madden20}. The surface density of molecular hydrogen (H$_2$), in particular, correlates with the SFR surface density, confirming that H$_2$ is the immediate fuel for star formation \citep{schmidt59,kennicutt12,bigiel08,leroy08,reyes19,sun23}. However, the cosmic H\,{\sc i} density declines more slowly than the H$_2$ density (see the left panel of Fig.\,\ref{fig:sf}), indicating that the conversion of H\,{\sc i} into H$_2$, rather than the raw supply of H\,{\sc i} itself, is the bottleneck for star formation \citep{walter20}. Understanding the atomic-to-molecular transition is therefore central to understanding star formation.
\\\indent The H$_2$ formation rate on dust grains, $R\sim3\times10^{-17}\,\mathrm{cm^3\,s^{-1}}$ \citep{hollenbach71,jura75a,jura75b}, together with the requirement for dense, self-shielding gas to protect H$_2$ against photodissociation \citep{krumholz09b,mckee10,sternberg14}, sets the physical requirements for molecular gas formation. Several prescriptions exist for the resulting molecular gas fraction $f_{\rm mol}$: an empirical pressure-based relation \citep{blitz06}
\begin{equation}
    f_{\rm mol}=\frac{(P_{\rm ext}/P_0)^\alpha}{1+(P_{\rm ext}/P_0)^\alpha}
,\end{equation}
where $P_{\rm ext}$ depends on the gas and stellar mass surface density and velocity dispersion, a first-principles formulation based on the UV radiation field ($U$) and dust-to-gas ratio ($D$) \citep{gnedin11}
\begin{equation}
    f_{\rm mol}=\frac{1}{1+e^{-4x-3x^3}}
,\end{equation}
where $x=f(D,U)$ is a function of $U$ and $D$, and a semi-analytical formulation based on the optical depth of H\,{\sc i} clouds \citep{krumholz13}
\begin{equation}
    f_{\rm mol}=1-\frac{3}{4}\frac{s}{1+0.25s}
,\end{equation}
where $s=f(Z,U)$ depends on the metallicity of the gas and dust phase ($Z$) and $U$. These approaches have yet to converge, and it remains unclear which, if any, correctly captures the underlying physics \citep{marasco25}.
\\\indent On galaxy-averaged scales, the SFR surface density correlates with the molecular gas surface density following a close-to-linear Kennicutt--Schmidt relation (\citep{schmidt59,kennicutt12,bigiel08,leroy08,reyes19,sun23}; see the right panel of Fig.\,\ref{fig:sf}). This relation is understood as the outcome of the gravitational collapse and fragmentation of giant molecular clouds (GMCs) into pre-stellar cores \citep{stahler04,mckee07,krumholz14}. The tightness of the correlation further indicates that the process is relatively rapid. Indeed, resolved, multiwavelength studies of nearby galaxies combining CO and H$\alpha$ tracers show that the lifetime of a GMC and the duration of the embedded star-formation episode it hosts are comparable and short, of order a few Myr \citep{chevance20}.

\begin{figure}[!t]
    \centering
    \includegraphics[width=\linewidth]{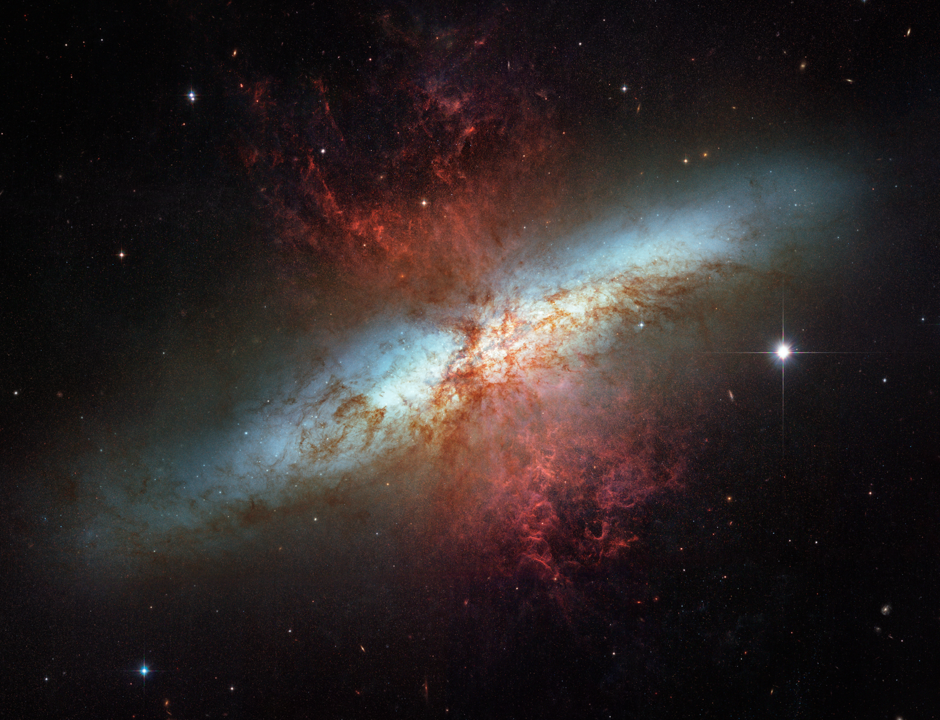}
    \caption{The galactic wind in M82. The background-coloured image is from HST, while the ionised gas emission observed by the Subaru Telescope is shown in red. Image credit: \href{https://hubblesite.org/contents/media/images/2006/14/1876-Image}{STScI}.}
    \label{fig:m82}
\end{figure}

\section{Stellar Feedback}\label{sec:stellarfeedback}
The impact of star formation on the baryon cycle is not limited to the depletion of gas reservoirs in galaxies. Massive ($>10$ M$_\odot$) stars exhibit significant mass ejection during their lifetime \citep{gehrz71,kudritzki00,vink22} and die in core-collapse supernovae \citep{janka07,burrows20,jerkstrand26}. The combined energy and momentum input from stellar winds and supernovae drives galaxy-scale outflows capable of ejecting gas from the ISM \citep{fierlinger16,zhang18}, as exemplified by the remarkable case of M82, shown in Fig.\,\ref{fig:m82}.
\\\indent Cosmological simulations predict that this stellar feedback shapes the galaxy luminosity function by suppressing star formation in low-mass ($M_\star\lesssim10^9\,M_\odot$) galaxies \citep{puchwein13,eagle,dave17}, explaining the deviation at the low-mass end of the observed galaxy mass function from the prediction of $\Lambda$CDM (e.g.\ \citep{baldry08,li09,silk12}). On galactic scales, stellar feedback regulates chemical evolution and contributes to the mass--metallicity relation \citep{tremonti04,finlator08,baldry08,peeples11,scholte24}. Galactic winds also heat the ISM, preventing the efficient formation of GMCs and thereby suppressing further star formation \citep{heckman00,bertone07}.
\\\indent However, stellar feedback also has a positive impact on star formation. Typical wind velocities ($<100$ km s$^{-1}$) are lower than the escape velocity of the galaxy. Consequently, the outflow does not escape the CGM and falls back into the galaxy mixed with fresh pristine CGM gas. This produces a galactic fountain that recycles gas and sustains future star formation \citep{fraternali13,fraternali17,chisholm16,voort17,krumholz17}.
\\\indent Despite the broad understanding of stellar feedback, many details remain unclear; for instance, how supernova explosions maintain gas turbulence in star-forming galaxies. Observations find that $\sim60\%$ of the atomic ISM resides at temperatures of 100 K and 8000 K \citep{kulkarni87,dickey90,dickey93}, while $\sim40\%$ is transitioning between these two phases \citep{heiles01}. Theoretical arguments likewise predict that the atomic ISM should be in thermal equilibrium at 100 K and 8000 K \citep{wolfire95}, corresponding to line dispersions of $\sim1$ km s$^{-1}$ and $\sim8$ km s$^{-1}$, respectively. The presence of turbulence in the gaseous disc is therefore empirically inferred from the detection of lines with dispersions $>8$ km s$^{-1}$. Such broader lines are typically found towards the galactic centre in both the atomic \citep{petric07,ianjam12,iorio17} and molecular \citep{wilson11,caldu13,mogotsi16} phases of the ISM. Potential origins of this turbulence include supernova feedback \citep{kim15b,martizzi16,fierlinger16,ohlin19}, magnetic \citep{dziour04,piontek05} and gravitational instabilities \citep{agertz09,bournaud10}, and accretion flows \citep{klessen10,krumholz10,elmegreen10}. Empirical studies \citep{tamburro09,stilp13,krumholz16,utomo19,bacchini20} have not yet provided a conclusive answer, being largely limited by the difficulty of estimating how much power feedback processes inject into the ISM relative to other sources of turbulence.

\begin{figure}[!t]
    \centering
    \includegraphics[width=\linewidth]{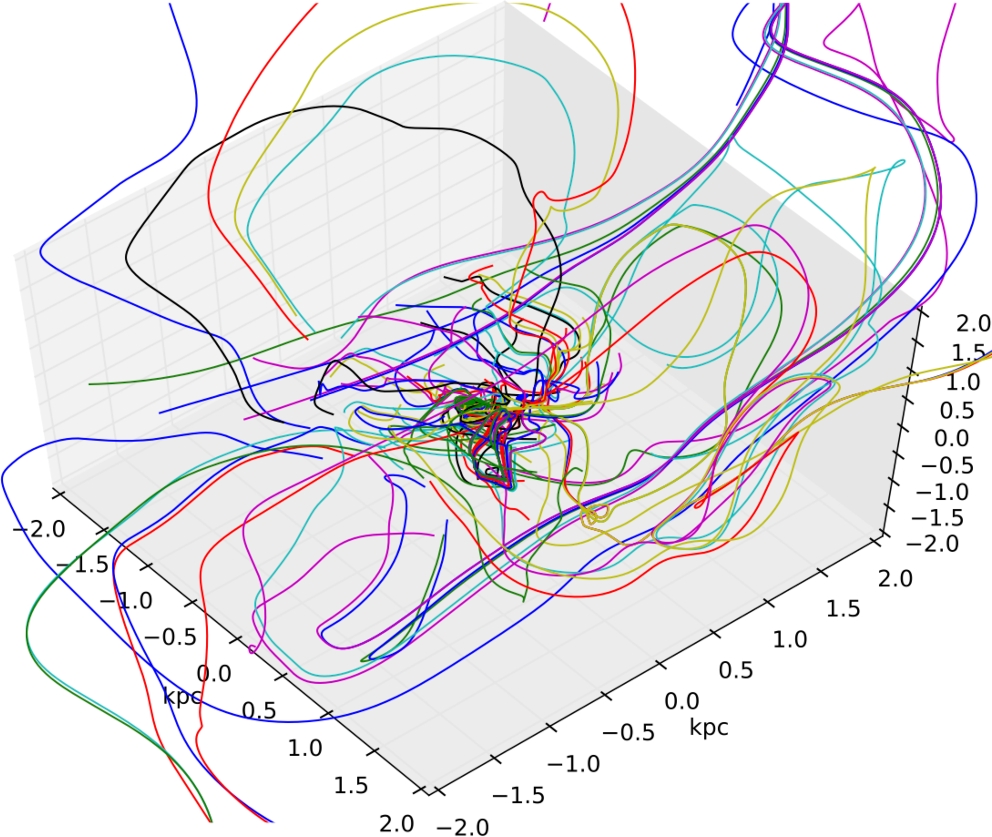}
    \caption{Sample of trajectories of CCA gas in the inner kpc region of a simulated galaxy \citep{gaspari17}.}
    \label{fig:cca}
\end{figure}

\section{AGN Feeding}\label{sec:agnfeeding}
The gas accreted from the cosmic web and the CGM does not entirely convert into stars. A fraction of the inflowing material feeds the SMBH hosted at the centre of almost every galaxy \citep{kormendy95,kormendy13,magorrian98,ferrarese00,gebhardt00}, including the Milky Way \citep{eht22b}. However, how the gas actually reaches the SMBH depends on cosmic epoch, halo mass, and host morphology, and several distinct feeding channels have accordingly been proposed.
\\\indent Within the Bondi radius of the SMBH (typically $r_{\rm B}\sim0.1$--300 pc), the classical Bondi solution \citep{bondi52} describes a local, spherically symmetric infall under the simplifying assumptions of negligible angular momentum and a steady density and temperature set at $r_{\rm B}$. Gas-rich mergers \citep{hopkins06,koss18,ellison19,ellison25,sharma24} and secular processes such as stellar bars \citep{galloway15,alonso18,smethurst19,silvalima22} and disc instabilities exert non-axisymmetric gravitational torques that remove angular momentum from circumnuclear gas, driving inflows from kpc down to pc scales. In hot, pressure-supported atmospheres typical of elliptical galaxies, groups, and clusters, thermal instabilities in the hot CGM can condense into cold clouds and filaments that decouple from the hot phase and rain inward under gravity and inelastic collisions, ultimately feeding the SMBH \citep{gaspari13,gaspari17,gaspari18,gaspari20}, a process known as chaotic cold accretion (CCA; see Fig.\,\ref{fig:cca}). CCA is closely related to the precipitation model, which links condensation to a roughly constant ratio of cooling to free-fall time \citep{voit15b,voit17}. The relevancy of CCA is that it ties cold-gas fuelling to a self-regulated AGN feedback loop that reheats the halo and resets the cycle, linking SMBH feeding and AGN feedback into a coherent framework.
\\\indent Observationally, CCA leaves a characteristic signature in the relationship between line width and line-of-sight velocity of the cold gas \citep{gaspari18}. The model has so far been tested mainly in elliptical galaxies \citep{maccagni18,maccagni21} and galaxy clusters \citep{voit15,storchi19}, leaving open the question of whether it applies to gas-rich systems such as disc galaxies, where rotational support and bar/spiral torques, rather than hot-halo condensation, are expected to dominate the inflow budget. Testing the CCA line width--velocity diagnostic in rotating, multiphase disc environments remains an outstanding observational challenge.

\begin{figure}[!t]
    \centering
    \includegraphics[width=\linewidth]{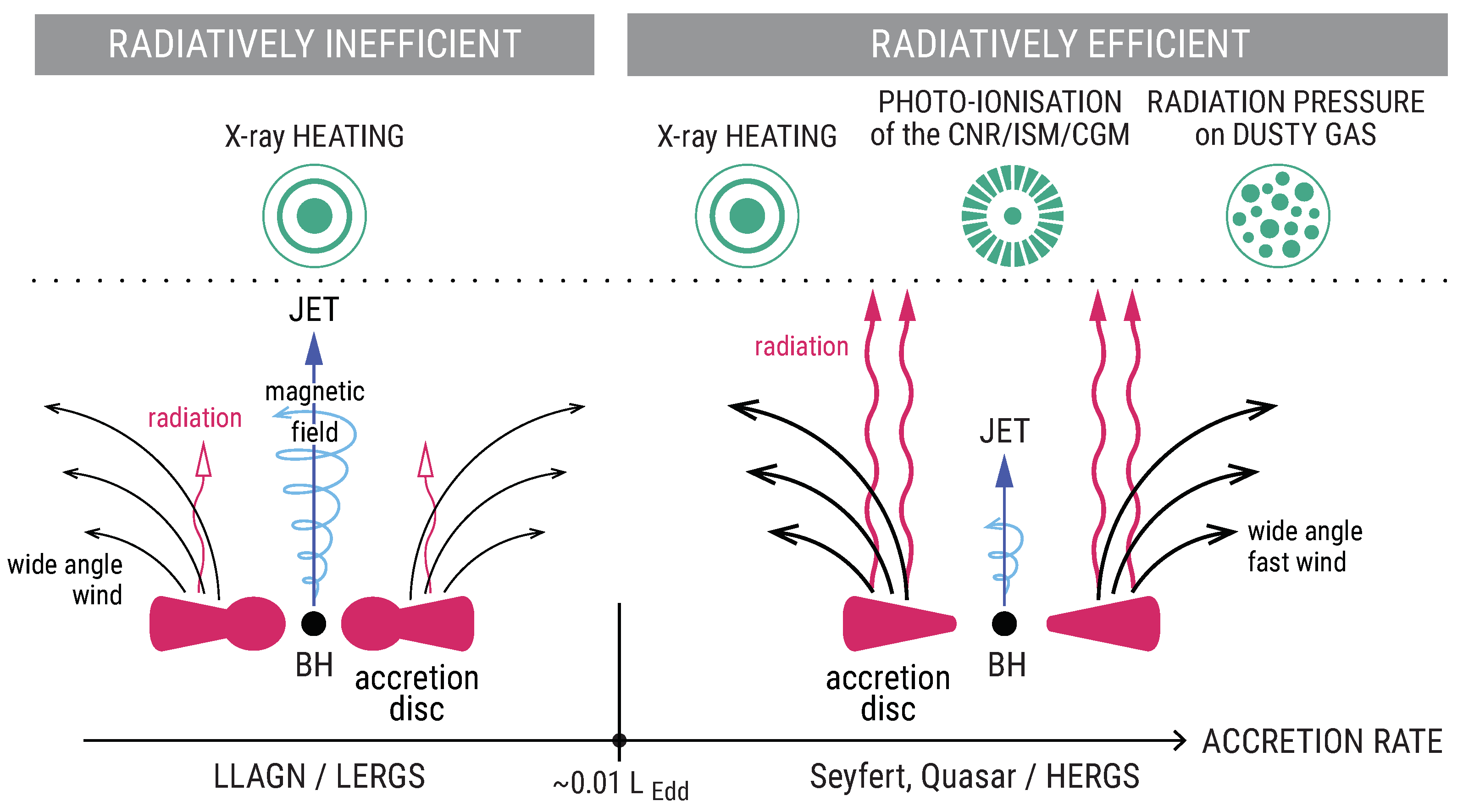}
    \caption{A schematic cartoon of the two feedback modes from AGN \citep{harrison24}. In the left panel, radiatively inefficient accretion onto the SMBH leads to the launch of powerful jets of relativistic particles. The main effect of the jets is to heat the gas on scales ranging from the ISM to the intergalactic medium. In the right panel, radiatively efficient accretion leads to the release of most of the energy in the form of radiation, inducing photoionisation of the surrounding medium and launching outflows via radiation pressure on dusty gas.}
    \label{fig:agn}
\end{figure}

\section{AGN Feedback}\label{sec:agnfeedback}
The feeding of the SMBH is directly related to the role of AGN feedback in regulating the growth of massive galaxies \citep{puchwein13,vogel14,heckman24,harrison24}. AGN feedback occurs in a radiative or a kinetic mode depending on how efficiently the SMBH is accreting material, as depicted schematically in Fig.\,\ref{fig:agn}.
\\\indent Radiative-mode AGN feedback is associated with radiatively efficient accretion onto the central SMBH \citep{shakura73}, often observed in luminous Seyfert galaxies. The high-energy radiation from the AGN photoionises the gas out to CGM scales \citep{arrigoni19,costa22} and drives strong outflows through radiation pressure \citep{ishibashi18,costa18} or thermal winds \citep{murray95,proga00,costa20,mizumoto21}. These outflows remove gas from the galaxy, suppressing star formation and regulating black hole growth \citep{dimatteo05,debuhr12,fiore17,torrey20,ricci22,massonneau23}. Observations of blue-shifted absorption \citep{wagner13,laurenti21,mizumoto21} and emission \citep{shih13,fiore17,molyneux19,santoro20,vietri20} lines in the optical, UV, and X-ray spectra of AGN, which can reach velocities of several thousand kilometres per second, provide direct evidence of these winds. The most extreme cases exhibit molecular and ionised outflows extending over kpc scales, capable of driving large reservoirs of gas out of the host galaxy \citep{feruglio10,cicone12,cicone14}.
\\\indent In contrast, kinetic AGN feedback, also known as maintenance-mode feedback, is primarily observed in higher-luminosity AGN hosted by massive elliptical galaxies \citep{sabater19}. This mode arises when the accretion flow onto the SMBH is radiatively inefficient and therefore releases most of its energy in the form of relativistic, highly collimated plasma ejections \citep{narayan94,yuan14}. These jets, extending from sub-kpc to Mpc scales (see the reviews by \citep{hardcastle20,odea21,dabhade23} and references therein), heat the surrounding gas, preventing it from cooling to form new stars. This mechanism is crucial for maintaining the red-and-dead nature of massive elliptical galaxies \citep{danielson12,dubois13,smolcic17,hardcastle19,su21,heckman24} and for regulating the growth of galaxy clusters \citep{mcnamara12,hlavacek22,ehlert23}. Observations of radio-loud AGN, particularly in brightest cluster galaxies, show that AGN-driven jets can inflate bubbles in the hot intracluster medium, providing a continuous source of heating that counteracts radiative cooling \citep{birzan12,gitti12,hlavacek12,hlavacek15,mcdonald18}. Recent observations of the low-luminosity AGN in the disc galaxy NGC 1371 \citep{veronese25b} reveal that even a weak jet is able to inflate bubbles and significantly suppress star formation at tens of kpc from the SMBH in gas-rich systems.
\\\indent Although AGN feedback has traditionally been considered a suppressive mechanism for star formation, and is included in cosmological simulations to reconcile the predicted properties of massive galaxies with observations by suppressing their growth \citep{eagle,khandai15,dubois16,mccarthy17,nelson18,simba}, recent studies suggest that, in some cases, AGN outflows and jets may also trigger star formation by compressing gas clouds \citep{ishibashi12,zinn13,silk13,cresci15a,cresci15b,gallagher19,zhuang20}. A comprehensive model of galaxy evolution must therefore integrate all feedback mechanisms, both stellar- and AGN-driven, within a unified framework describing their interplay. Some studies have already begun to tackle this effort \citep{biernacki18,li18}.

\section{Conclusion and future prospects}\label{sec:conc}
This review has outlined how the baryon cycle links the cosmic web, the CGM, and the ISM of galaxies through accretion, star formation, and feedback. While the broad framework is well established, several important questions remain open at each stage of the cycle.
\\\indent Gas accretion onto galaxies is still not directly observed in an unambiguous way in the nearby Universe. Although cold-mode accretion from the CGM is thought to dominate gas supply at low redshift, as it did at early times \citep{correa18,waterval25}, the measured accretion rate at $z\sim0$ does not match the observed star formation rate \citep{sancisi08,putman12,diteodoro14,kamphuis22}, and the kinematically anomalous H\,{\sc i} gas seen around nearby galaxies has no firmly established origin \citep{fraternali02,braun04,pisano14,oosterloo07,wolfe16,xu22,liu23,das24}. Deeper and wider H\,{\sc i} surveys, together with more sensitive tracers of diffuse, ionised gas, will be needed to close this gap \citep{mhongoose2,veronese25,yang25,wang26}.
\\\indent Within the ISM, the atomic-to-molecular transition remains the key bottleneck controlling where and when star formation occurs \citep{walter20}. The different theoretical approaches to this transition, such as pressure-based \citep{blitz06}, radiation- and dust-based \citep{gnedin11}, and opacity-based \citep{krumholz13}, have yet to converge on a single, consistent picture \citep{marasco25}. Similarly, although the Kennicutt--Schmidt relation \citep{sun23} and the short lifetimes of giant molecular clouds \citep{chevance20} point to a rapid, localised star-formation process, the sources that sustain turbulence in the atomic ISM, like supernovae, magnetic or gravitational instabilities, or accretion flows, remain difficult to disentangle observationally \citep{tamburro09,stilp13,krumholz16,utomo19,bacchini20}.
\\\indent Stellar feedback plays a dual role, suppressing star formation in low-mass galaxies \citep{puchwein13,eagle,dave17} while also recycling gas through galactic fountains that can sustain future star formation \citep{fraternali13,fraternali17}. The balance between these competing effects, and how it varies with galaxy mass and environment, is not yet fully quantified.
\\\indent At galactic centres, several channels for feeding SMBHs have been proposed, from classical Bondi accretion \citep{bondi52} to merger- and bar-driven inflows \citep{hopkins06,galloway15} and cold chaotic accretion from a hot, condensing CGM \citep{gaspari13,gaspari17,gaspari18,gaspari20}. The latter, able to naturally link SMBH feeding with AGN feedback loop, has so far been tested mainly in elliptical galaxies \citep{maccagni18,maccagni21} and galaxy clusters \citep{voit15,storchi19}, and whether it also operates in rotationally supported disc galaxies remains an open observational challenge.
\\\indent Finally, AGN feedback is traditionally modelled as a purely suppressive process in cosmological simulations \citep{eagle,khandai15,dubois16,mccarthy17,nelson18,simba}, yet growing observational evidence shows that AGN outflows and jets can also trigger star formation by compressing gas \citep{ishibashi12,zinn13,silk13,cresci15a,cresci15b,gallagher19,zhuang20}. Reconciling these apparently opposing effects, and combining them with stellar feedback into a single, self-consistent framework, is an active area of research \citep{biernacki18,li18}.
\\\indent In the near future, a number of upcoming facilities will open new observational windows on the Universe across the entire electromagnetic spectrum. At low frequencies (50 MHz--350 MHz), the radio telescope SKA-Low in Australia will enable precise broad-band spectral index measurements, necessary to identify AGN and characterise their age in order to learn how low-power jets expand through the ISM and CGM. SKA-Mid, located in South Africa, covers the 0.35--15.4 GHz band and, with $\sim100$ hours per pointing, can potentially reveal, for the first time, H\,{\sc i} in the CGM associated with cold accretion in the local Universe. In the northern hemisphere, the future Deep Synoptic Array (DSA; \citep{dsa2000}) can be used to identify many more low-luminosity AGN in disc galaxies. DSA will also map the 1.4 GHz continuum sky ($\sim31000$ deg$^{2}$) at an unprecedented angular resolution of $\sim3''$. The planned Next Generation VLA observatory (ngVLA; \citep{ngvla}) can then be used for targeted follow-up studies.
\\\indent The upcoming Legacy Survey of Space and Time (LSST; \citep{lsst}) on the Vera Rubin Observatory \citep{rubin} will provide ultra-deep optical observations ($5\sigma$ limiting magnitudes ranging from 24.9 mag in the $y$ filter to 27.5 mag in the $r$ filter) of $\sim18000$ deg$^{2}$ of the southern sky at $\sim1''$ resolution. Deep optical surveys of the sky are concurrently being performed by the Euclid space telescope \citep{euclidb}. The Euclid Wide Survey (EWS; \citep{ews}) is acquiring $\sim0.2''$-resolution images in the 530--920 nm band over $\sim14500$ deg$^{2}$ of the extragalactic sky. Such optical surveys represent the baseline catalogue for sample selection at other wavelengths.
\\\indent The hot ($T\gtrsim10^6$ K) gas that dominates the volume of the CGM, and which is fundamental to testing hot-mode accretion, can be probed in emission only with X-ray observatories, and has so far been detected around only a handful of luminous galaxies. The Advanced Telescope for High ENergy Astrophysics (ATHENA; \citep{athena}), planned for launch in the late 2030s, will represent a major improvement over current X-ray instruments such as Chandra \citep{chandra}, XMM-Newton \citep{xmm}, and the X-Ray Imaging and Spectroscopy Mission (XRISM; \citep{xrism}). It will offer a comparable angular resolution ($\sim5''$), energy range (0.2--12 keV), and spectral resolution ($<4$ eV at 7 keV) to these facilities, together with a much larger field of view ($\sim40'$) and an effective area $\sim1.5$ dex higher, i.e.\ 5--100 times more sensitive, depending on the energy considered. ATHENA will thus provide unprecedented opportunities to characterise the hot CGM of nearby galaxies.

\bibliographystyle{JHEP}
\bibliography{references.bib}

\end{document}